# Demonstration of a High-$Q$ Subwavelength Dielectric Nanocylinder

DAYANG LI[1,2,†], SIMON KLINCK BORREGAARD[1,2,†], JESPER MØRK[1,2], MENG XIONG[1,2], AND YI YU[1,2,*]

[1]*Department of Electrical and Photonics Engineering, Technical University of Denmark, Ørsteds Plads 345A, Kgs. Lyngby 2800, Denmark*
[2]*NanoPhoton - Center for Nanophotonics, Ørsteds Plads 345A, Kgs. Lyngby 2800, Denmark*
[†]*These authors contributed equally to this work.*
[*]*yiyu@dtu.dk*

## 1. Abstract

The development of subwavelength dielectric cavities is essential for reducing the size of photonic devices and enabling dense optoelectronic integration. However, previously subwavelength optical cavities exhibit demonstrated $Q$-factors <400, limiting their applications. Here, we demonstrate a high-$Q$ subwavelength nanocylinder by leveraging bound states in the continuum (BIC). We track BIC modes of different longitudinal orders while maintaining an ultrasmall footprint. We find that the $Q$-factor initially increases but then saturates at higher orders. By linking quasi-normal-mode perturbation theory with coupled-mode analysis, we reveal the physical origin of this saturation and identify an optimized dimension that balances performance with fabrication feasibility. By suspending this design in free space using nanobridges and optimizing the nanofabrication process, we experimentally realize an InP subwavelength nanocylinder with a measured $Q$-factor exceeding 1000. Compared with a lower-$Q$ substrate-supported counterpart, the suspended high-$Q$ BIC nanocylinder exhibits stronger scattering and photoluminescence signals. Our work provides a route to high-$Q$ optical devices with ultrasmall footprints.

## 2. Introduction

The continued scaling of electronic integrated circuits has created a strong demand for optical and optoelectronic components with increasingly small footprints [1–4]. Although remarkable progress has been made in the realization of ultrahigh quality-factor ($Q$-factor) cavities [5–10] and ultrasmall mode volumes [11–19], the physical footprints of photonic devices typically remain orders of magnitude larger than those of their electronic counterparts. Therefore, developing efficient optical cavities with ultrasmall, ideally subwavelength, footprints is essential for bridging the gap between photonics and electronics, while also enhancing light-matter interactions, lowering energy consumption, and enabling highly compact optoelectronic systems.

Subwavelength dielectric cavities provide a promising route because they can confine light with relatively low material loss compared with plasmonic structures [20–22]. Nevertheless, achieving a high $Q$-factor in such open systems remains challenging due to the strong radiation leakage. This has motivated the use of bound-states-in-the-continuum (BICs) [23], particularly Friedrich-Wintgen BICs [24–30], where destructive interference between two leaky modes—typically Fabry-Perot (FP)-like and Mie-like modes within the same compact cavity—suppresses far-field radiation and gives rise to a high-$Q$ supercavity mode or quasi-BIC (referred here simply as the BIC), accompanied by a complementary, more strongly radiating branch (referred here as the anti-BIC (A-BIC)). However, experimental demonstrations of such optical BICs have so far remained limited, with reported $Q$-factors still below 400 [27–29].

Here, we demonstrate a BIC with a $Q$-factor exceeding 1000 in a subwavelength dielectric nanocylinder operating at telecom wavelengths. We further find that the BIC $Q$-factor initially increases with structure height but subsequently saturates. By developing coupled-mode models mapped from quasi-normal-mode perturbation theory, we provide insight into the physical origin of this saturation.

## 3. Theoretical analysis

For a nanocylinder with radius $r$ and height $h$ (inset of Fig. 1(a)), we consider rotationally symmetric TE modes, which can be generally approximated in cylindrical coordinates $(\rho, \phi, z)$ as $E(\mathbf{r}) = J_1(x_n\rho/r)e^{jl\phi}(A\sin(m\pi z/h) + B\cos(m\pi z/h))$ [31], where $J_1$ is the first-order Bessel function of the first kind, and $x_n$ is its $n$'th zero. $l, m$ and $n$ denote the azimuthal ($\phi$), longitudinal ($z$) and radial ($\rho$) mode orders, respectively. We focus on ultracompact devices and therefore restrict our analysis to the lowest azimuthal and radial order ($l = 0$ and $n = 1$), while varying $m$ by tuning $h$ (or the aspect ratio $R = h/r$). The corresponding eigenwavelength is given by $\lambda_{m,n} = 2\pi n_{\mathbf{e}}/\sqrt{(m\pi/h)^2 + (x_n/r)^2}$ (see SI, Section 1). For clarity, we label each mode by $(n_z, n_r)$, where $n_z$ ($n_z = m$) and $n_r$ ($n_r = 2n$) denote the numbers of antinodes in $|\mathbf{E}|$ along the $z$ and lateral directions, respectively. Friedrich-Wintgen BICs are typically accompanied by avoided crossings, arising from intersections between originally uncoupled mode branches. We therefore consider two branches whose frequencies can cross: an FP-like mode with indices (O+2, 2), and a Mie-like mode with indices (O, 4), where O=1, 2, 3. . . .

Using $\lambda_{m,n}$, six mode pairs are tracked by varying $h$, cf. Fig. 1(a) (here $r$=0.56 μm). Multiple crossings arise between the FP-like modes (solid curves), whose resonant frequencies vary strongly with $h$, and the Mie-like modes (dashed curves), whose resonant frequencies depend weakly on $h$. Here, we only consider crossings between (O+2, 2) and (O, 4), since only modes with the same symmetry, i.e., both having even or odd indices, can interact, and these crossings occur at relatively small $R$, which is advantageous for nanofabrication. Assuming that the effective index $n_\mathrm{e}$ is identical for the two mode branches, $h$ at the crossing point ($h_\mathrm{O}$) can be obtained (see SI, Section 1) as

$$h_\mathrm{O} \approx 2\pi r\sqrt{(\mathrm{O}+1)/\left(x_2^2 - x_1^2\right)} = 1.07r\sqrt{\mathrm{O}+1}\,. \tag{1}$$

The corresponding resonant wavelength is given by

$$\lambda_\mathrm{O} = 4\pi n_\mathrm{e} r\sqrt{\frac{\mathrm{O}+1}{(\mathrm{O}+2)^2 x_2^2 - \mathrm{O}^2 x_1^2}} \approx \frac{2n_\mathrm{e} r}{\sqrt{\mathrm{O}}} \qquad \text{(for large O)}, \tag{2}$$

It should be noted that Eqs. (1) and (2) are derived under lossless hard-boundary conditions and therefore cannot account for radiative coupling. Nevertheless, they provide useful estimate for BIC mode tracking. We verify this using finite-element-method (FEM) simulations with perfectly matched layers to account for the open nature of the system. Specifically, we consider an InP nanocylinder with a constant refractive index of 3.17 and use $h_\mathrm{O}$ and $\lambda_\mathrm{O}$ as initial conditions to identify the BIC locations. Fig. 2(a) shows that, under open-boundary conditions, the original crossings evolve into avoided crossings occurring close to $h_\mathrm{O}$, and $n_\mathrm{e}$ is extracted to be 3.6~4. This confirms the validity of the analytical estimates and indicates that the modes interact through radiation channels when they are close in frequency. The field profiles of the BIC mode (Fig. 1(b)) inherit the same symmetries of the FP- and Mie-like modes. Specifically, the BIC mode shares the logitudinal (transverse) antinode count of the Mie (FP)-like mode, reflecting their hybridization. It is therefore indexed as $(\mathrm{O}, 2)$, where O denotes its order.

Around each avoided crossing, the outward radiation of the BIC branch is strongly suppressed, maximizing its $Q$-factor, while the A-BIC branch's $Q$ reaches a minimum. Notably, the $Q$-values of the optimum BIC states increase initially but saturate at approximately 1000 for O$\geq$ 3 (Fig. 2(b)). To understand this behavior, we performed a quasi-normal-mode perturbation theory (PT) analysis [32], which shows good agreement with the full numerical results (Figs. 2(a,b)).

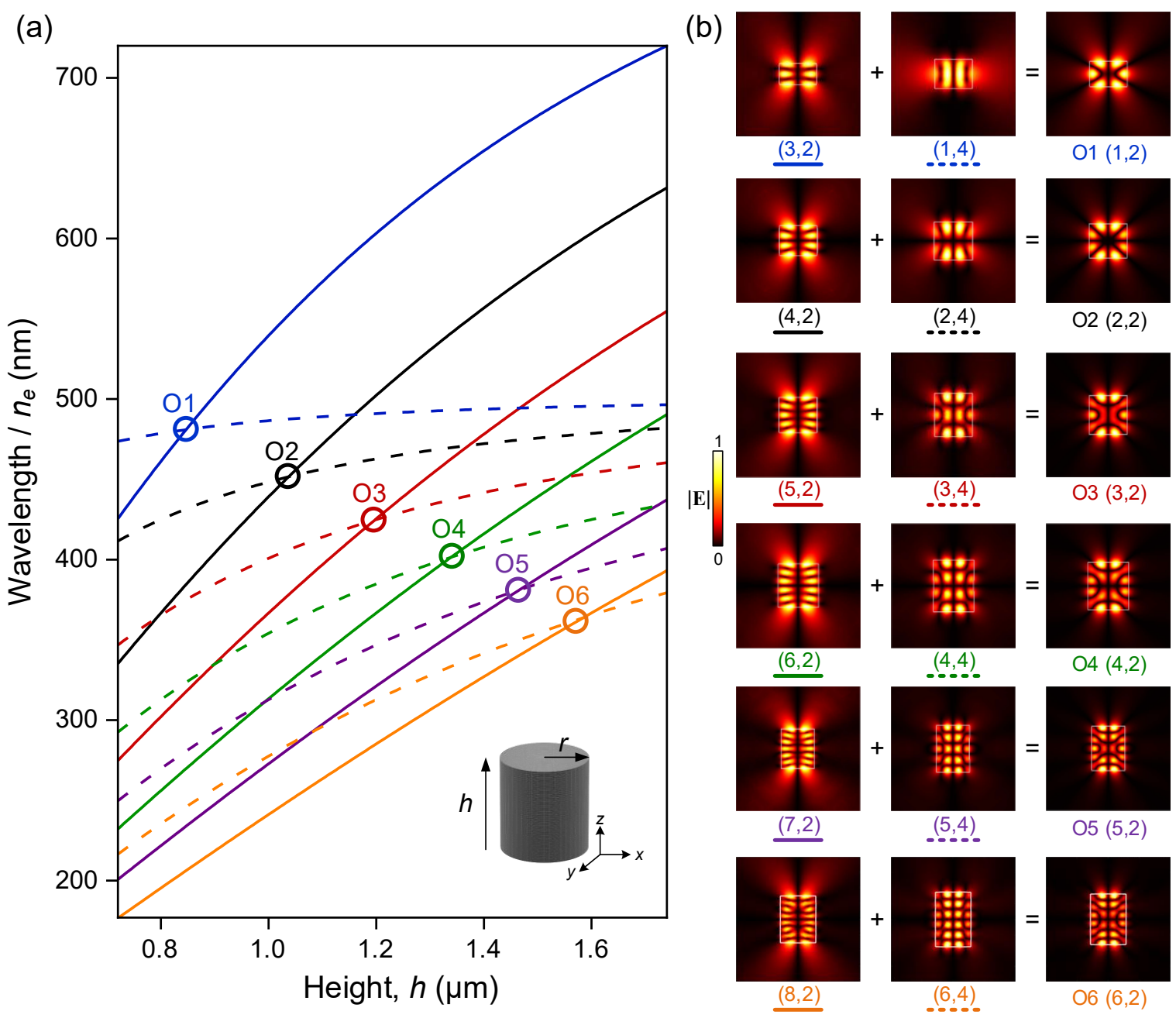

Fig. 1. (a) Normalized resonant wavelength evolution of the Fabry-Perot (FP) like modes (solid curves) and Mie-like modes (dashed curves) as the height $h$ of a nanocylinder (inset) varies. The intersection points between the FP- and Mie-like branches are denoted O1-O6. (b) Corresponding electric-field magnitude distributions ($|\mathbf{E}|$) of the modes in the $x$-$z$ cross-section, obtained from finite-element (FEM) simulations, showing the FP-like modes (left column), the Mie-like modes (middle column), and the resulting quasi-bound-state-in-the-continuum (BIC) modes (right column) at the intersection points, arranged from the lower-order case O1 (top) to the higher-order case O6 (bottom). The white rectangles outline the nanocylinder boundaries. The modes are labeled as $(n_z, n_r)$, where $n_z$ denotes the number of $|\mathbf{E}|$ antinodes along the height direction and $n_r$ denotes the number of $|\mathbf{E}|$ antinodes along the lateral direction.

Specifically, we consider nanocylinders whose heights deviate by $\Delta h$ from $h_\mathrm{B}$ ($h_\mathrm{B} \approx h_\mathrm{O}$), the height at which the BIC branch reaches its maximum $Q$, i.e., the structures must decrease or increase in height to reach $h_\mathrm{B}$, as illustrated in the insets of Fig. 2(d). For these perturbations, representing the system as a standard eigenvalue problem yields an effective perturbative Hamiltonian (see SI, Section 2): $\mathbf{H}_\mathrm{PT} = [\mathbf{I} + \mathbf{H}_\mathrm{P}]^{-1}\mathbf{H}_0$, where $\mathbf{H}_0$ is a diagonal matrix with elements being complex eigenfrequencies of the unperturbed modes, $\mathbf{I}$ is the identity matrix, and $\mathbf{H}_\mathrm{P}$ is the perturbation matrix arising from geometric deformation, whose elements, under the first-order approximation, can be evaluated via a surface integral

$$\mathbf{H}_{\mathrm{P},uv} = \int_{\partial V} \mathbf{E}_u(\mathbf{r}_{\partial V} - \sigma\hat{\mathbf{n}}) \cdot \Delta h(\mathbf{r}_{\partial V})\Delta\varepsilon \cdot \mathbf{E}_v(\mathbf{r}_{\partial V} + \sigma\hat{\mathbf{n}})\mathrm{d}^2\mathbf{r}_{\partial V}, \tag{3}$$

where $\mathbf{E}_u(\mathbf{r}_{\partial V} - \sigma\hat{\mathbf{n}})$ and $\mathbf{E}_u(\mathbf{r}_{\partial V} + \sigma\hat{\mathbf{n}})$ ($\sigma \to 0$) are the normalized complex electric fields of the unperturbed modes $u$ and $v$, evaluated at infinitesimally close points on the inner and outer sides of the object surface at position $\mathbf{r}_{\partial V}$; $\hat{\mathbf{n}}$ denotes the outward unit normal vector at $\mathbf{r}_{\partial V}$; $\Delta h(\mathbf{r}_{\partial V})$ is the boundary shift, and $\Delta\varepsilon$ is the permittivity difference between the object and the surrounding material (air in our case). In $\mathbf{H}_\mathrm{PT}$ the off-diagonal term $\eta = \eta_r + j\eta_i$ reflects the mode coupling. It remains nearly zero for most mode pairs, except between the FP- and Mie-like branches, mathematically confirming the onset of their interaction and the interference origin of the accidental BIC [23]. However, $\mathbf{H}_{\mathrm{PT}}$ itself cannot be interpreted as a standard coupled-mode matrix and therefore does not have a direct physical interpretation. Its elements vary strongly with $\Delta h$ (Fig. 2(c, I)), and its diagonal terms can even exhibit positive imaginary parts, corresponding to negative $Q$-factors and thus an unphysical effective gain.

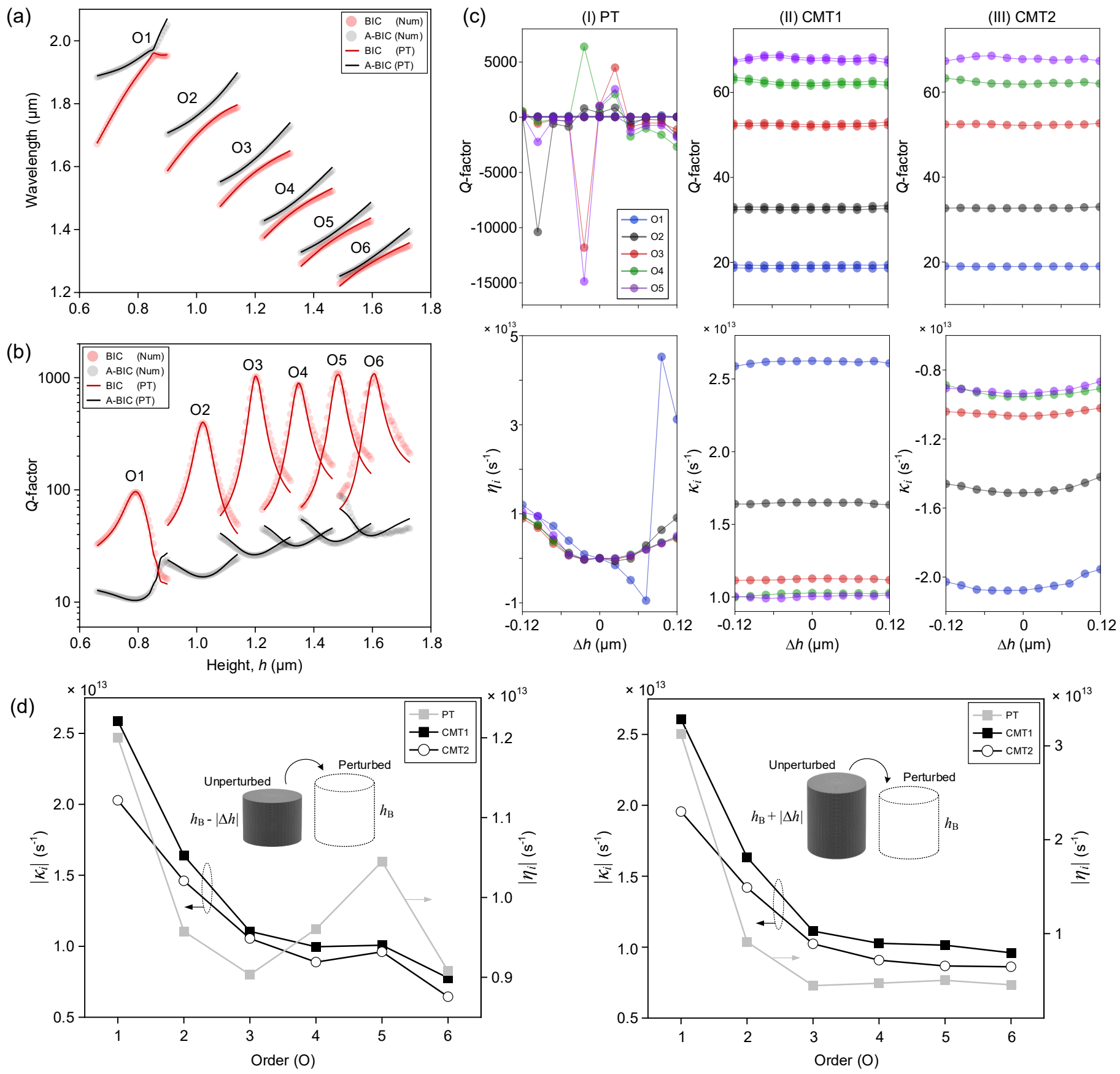

Fig. 2. (a,b) Variation of (a) wavelength and (b) $Q$-factor of the BIC (red) and A-BIC (gray) modes as a function of the nanocylinder height for mode order O=1–6. The colored dots are full numerical FEM results, while the solid lines represent semi-analytical predictions from quasi-normal-mode perturbation theory (PT), using the maximum-BIC-$Q$ point of each order as the unperturbed reference. (c) $Q$-factor, derived from the ratio between the real and imaginary parts of the diagonal elements (upper panels), and imaginary coupling components, $\eta_i$ and $\kappa_i$, extracted from the off-diagonal elements (lower panels). These quantities are obtained directly from the PT-based effective Hamiltonian $\mathbf{H}_{\mathrm{PT}}$ (I, left), and from two mapped coupled-mode matrices $\mathbf{H}_{\mathrm{CMT}}$ describing the coupling between the FP- and Mie-like modes: CMT1 (II, middle) and CMT2 (III, right). The quantities are plotted as a function of the perturbation, defined as a cylinder height deviation $\Delta h$ from the optimal one $h_{\mathrm{B}}$, where the BIC $Q$ is maximized for each order. The different colors correspond to different O, ranging from 1 to 5. (d) $\kappa_i$ ($\eta_i$), shown in black (gray), as a function of O. The insets illustrate that the coupling is induced when the initial unperturbed nanocylinder height is $\Delta h$ = 0.12 µm smaller than (left) or larger than (right) $h_{\mathrm{B}}$ in the perturbed case. Here, a relatively large $\Delta h$ is chosen so that the unperturbed modes remain close to the original mode branches, allowing their individual properties to be used to predict the BIC optimum, while still remaining within the range where PT agrees well with the exact solutions.

To gain a clear physical insight, we reformulate $\mathbf{H}_{\mathrm{PT}}$ using a standard dual-mode coupled-mode theory (CMT) [33–35] (see SI, Sections 2 and 3)

$$\mathbf{H}_{\mathrm{CMT}} = \begin{bmatrix} \omega_1 - j\gamma_1 & \kappa_r + j\kappa_i \\ \kappa_r + j\kappa_i & \omega_2 - j\gamma_2 \end{bmatrix}, \tag{4}$$

where $\omega_i$ and $\gamma_i$ ($i = 1, 2$) denote the resonant frequencies and decay rates of the FP- and Mie-like modes projected onto a new basis, while $\kappa = \kappa_r + j\kappa_i$ represents their mutual coupling coefficient. Here, we focus on $\kappa_i$, as it governs the modal loss of the eigenmodes, whereas $\kappa_r$ primarily determines the frequency splitting (see SI, Section 3). Since the mapping from $\mathbf{H}_{\mathrm{PT}}$ to $\mathbf{H}_{\mathrm{CMT}}$ is non-unique, we consider two coupled systems commonly used for the physical interpretation of Friedrich-Wintgen BICs, in which effective interference arises when two modes have similar far-field radiation patterns but are out of phase. In the first model (CMT1) we impose $\gamma_1 = \gamma_2 = \gamma$ and $|\kappa_i| = \sqrt{\gamma_1\gamma_2} = \gamma$, meaning that the uncoupled modes have identical loss rates and that the loss coupling arises entirely from their shared radiative channels. In the second model (CMT2) we impose $\gamma_1 = \gamma_2 = \gamma$ and $\omega_1 = \omega_2 = \omega_0$, meaning that the uncoupled modes are degenerate.

As shown in Fig. 2(c, II and III), the mapped $Q$-factors ($\omega_i/(2\gamma_i)$) and coupling coefficients obtained from CMT1 and CMT2 are very similar, and they depend weakly on $\Delta h$ and exhibit clear trends with O, e.g., as O increases, $|\kappa_i|$ initially decreases rapidly and then tends to saturate for O$\geq$ 3 (Fig. 2(d)). We note that the value of $|\kappa_i|$ for O= 6 is less accurate, because the two-mode model becomes less reliable for larger structures, where interactions with additional modes are no longer negligible. Therefore, the O=6 results are not plotted in Fig. 2(c). We also examined other values of $\Delta h$ and cases with fixed $\pm\Delta h/h_{\mathrm{B}}$, and found that the saturation of $|\kappa_i|$ persists. This result is consistent with the observed $Q$ saturation. More specifically, the $Q$-factor is determined by $|\kappa_i|\ (\Delta\omega/|\kappa|)^2$ and $\gamma - |\kappa_i|$ in the CMT1 and CMT2 cases, respectively (see SI, Section 3), both of which are found to also saturate with O. We attribute this behavior to the reduced variation of the modal patterns with increasing $h$. As seen in Fig. 1(b), once $h$ exceeds the O3 case, the radiation pattern of both the FP- and Mie-like modes, particularly above and below the top and bottom surfaces of the nanocylinder, shows negligible variation. This weak evolution therefore leads to a much slower increase in the $Q$-factor at higher O. This also raises an interesting question regarding the maximum achievable BIC-$Q$ in such a structure. For example, by revealing the geometric dependence of $\gamma$ and $\kappa_i$, this analysis may provide routes to deterministically tune these parameters and optimize their ratio, thereby maximizing $Q$—an aspect that will be explored in future work.

In the present study, Fig. 2 indicates that the O3 BIC offers an optimal balance, achieving a high $Q$ (>1000) with moderate dimensions ($h$ or $R$), which is favorable for nanofabrication. In addition, this optimum $Q$ is the highest within a wavelength range as broad as 200 nm. In contrast, for O≥4, whispering-gallery modes with higher $Q$-factors and weaker dependence on geometric parameters appear nearby (see SI, Section 4), making the BIC harder to distinguish and compromising its advantage.

## 4. Experimental realization

We now focus on the realization of the O3 nanocylinder. The previous simulations consider nanocylinders in free space. However, in practice, the nanocylinder must either be placed on a substrate or suspended by supporting structures. Placement on a substrate typically leads to a significant decrease in the $Q$-factor, e.g., to ~300 on $SiO_2$ due to enhanced downward leakage (Fig. 3(a)). Here, we propose a design supported by nanobridges, cf. Fig. 3(c). The bridges are arranged with fourfold rotational symmetry, with each bridge having the same height as the nanocylinder at 1.2 μm and a width of 129 nm. Simulations show that these bridges provide enough mechanical strength while remaining sufficiently narrow to avoid strong perturbation of the modal field (see SI, Section 5). Interestingly, we found that an optimized bridge configuration can even in some cases lead to a slight enhancement in $Q$.

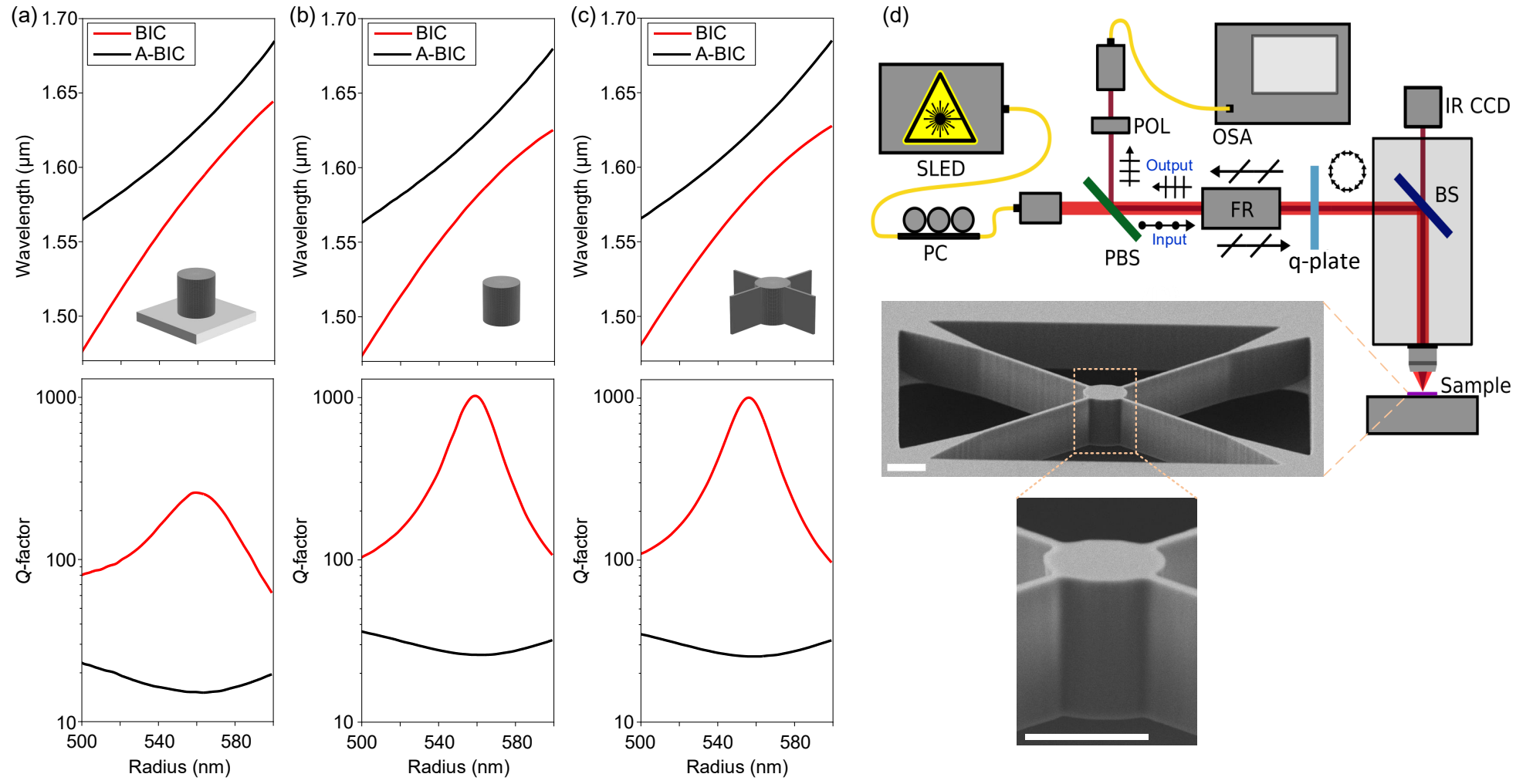

Fig. 3. (a-c) Three configurations of an InP O3 nanocylinder with a height of 1.2 μm (insets): (a) a nanocylinder on $SiO_2$ substrate, (b) a nanocylinder in free space, and (c) a nanocylinder suspended in free space via thin bridges (our experimental configuration). The top and bottom panels show the resonant wavelengths and $Q$-factors of the BIC (red) and A-BIC (black) mode, respectively, versus the nanocylinder radius. (d) Schematic of the scattering measurement setup and corresponding device morphology. The bottom panels show scanning electron microscope (SEM) image of a fabricated InP nanocylinder, together with a zoomed-in view. Scale bar: 1 μm.

Although the nanocylinder geometry is simple, its fabrication is not straightforward. For high-$Q$ subwavelength cavities, the key challenge is not only achieving a large etch depth, but also maintaining vertical and smooth InP sidewalls, since surface roughening, tapering, and notching [36] can perturb the cavity symmetry and introduce additional scattering loss, which is particularly detrimental for BIC modes that rely on destructive interference between leaky resonances. To limit mask roughening and maintain anisotropic etching before significant lateral erosion occurs, we use a relatively high InP etch rate of  8 nm/s. This places stringent demands on mask durability and pattern fidelity, requiring a careful optimization of both the lithographic resist and hard-mask thicknesses. Specifically, the fabrication proceeds as follows. A 1.2 μm epitaxial InP layer bonded onto an approximately 2.7 μm $SiO_2$ layer on a Si substrate is first coated with a 200 nm $SiN_x$ hard mask, followed by spin-coating a 500 nm electron-beam resist. The pattern is defined using a 100 kV JEOL JBX-9500FS electron-beam lithography system at a dose of 260 $\mu C/cm^2$, developed in N-50 for 3 minutes, and transferred into the hard mask via inductively coupled plasma (ICP) etching using a $CF_4$-based recipe. After resist stripping and the oxygen plasma cleaning, the pattern is subsequently transferred into the InP using a $Cl_2/H_2$ ICP reactive-ion etching process [37]. Finally, the structures are released through 30 minutes of BHF membranization, leaving the nanocylinders suspended by nanobridges. The optimized process enables high-fidelity transfer of features as small as 30 nm through the full 1.2 μm InP layer, corresponding to aspect ratios of up to 40 : 1, while preserving smooth, anisotropic, and vertical sidewalls.

The samples are characterized using scattering measurements (Fig. 3(d)). Since the BIC modes are rotationally symmetric and dominated by the $\phi$-field component, the setup is designed to generate azimuthally polarized light in order to maximize the coupling between the incident beam and the BIC. Specifically, a broadband source (SLED, Exalos EXS210066-02, total spectral power under objective -8dBm) is coupled through a polarization controller (PC) to prepare the input light in a linear polarization, allowing it to pass through the polarization beam splitter (PBS). The polarization is then rotated by 45° using a Faraday rotator (FR), after which the beam passes through a q-plate and is converted into either an azimuthally or radially polarized beam. The light is subsequently reflected downward by a beam splitter (BS) and focused onto the sample through an objective lens. The reflected (scattered) light from the sample is partially transmitted through the BS and monitored by an infrared camera. The remaining portion is reflected by the BS and passes back through the q-plate and the FR, after which it is converted back into linearly polarized output light with a polarization orthogonal to that of the input beam. It is therefore reflected by the PBS, passes through a polarizer (POL), and is finally collected and analyzed by an optical spectrum analyzer (OSA).

Fig. 4(a) shows scattering spectra measured for nanocylinders with fabricated radii ranging from 500 to 600 nm. The BIC mode is clearly resolved and shows excellent agreement with numerical simulations. As a sanity check, measurements under radially polarized excitation were also performed (see SI, Section 6), where the BIC mode is hardly excited due to polarization mismatch, consistent with simulations. The A-BIC mode, predicted by simulations and shown by the black line, is difficult to identify experimentally because of its weak scattering intensity, broad linewidth comparable to the slowly varying background, and proximity to the SLED long-wavelength cutoff around 1.62 μm. This weak A-BIC signature is also consistent with our finite-difference time-domain simulations. Although the A-BIC is difficult to resolve, a nearby whispering-gallery "crossover" mode, which passes through the BIC branch without an avoided crossing, is observed and agrees well with simulations, serving as a good reference.

The lower panels in Fig. 4(a) presents three representative spectra extracted across the optimum BIC point. A clear Fano-shaped resonance is observed, whose linewidth first narrows and then broadens, reflecting the strong geometric dependence characteristic of an accidental BIC. For comparison, we also measured the scattering spectra of nanocylinders on a $SiO_2$ substrate, without membrane release (Fig. 4(b)). In this case, the collected power at the optimum BIC is reduced from ~9.3 nW for the suspended case to ~0.72 nW for the substrate-supported structure, corresponding to an ~13-fold reduction. This reduction mainly arises from the increased fraction of optical leakage into the substrate, which decreases the upward radiation collected by the objective. The measured ratio agrees well with the CMT expectation of $(Q_{\mathrm{sus}}/Q_{\mathrm{sub}})^2$, where $Q_{\mathrm{sus}}$ and $Q_{\mathrm{sub}}$ are the BIC $Q$-factors of the suspended and substrate-supported structures, respectively. By fitting the spectra with a Fano lineshape, as shown by the red curves in Fig. 4(a,b), we extract the resonant wavelength and $Q$-factor. We note that the detailed Fano lineshape is sensitive to the excitation and collection conditions, we therefore performed multiple measurements to ensure the reliability of the extracted $Q$-factors. We also note that the BIC mode becomes less visible as the nanocylinder radius increases, as indicated by the reduced contrast of spectra III, therefore, the measured $Q$-factors are not plotted for all radii. The experiments agree very well with the simulations. For the suspended nanocylinders, a maximum $Q$-factor of $Q_{\mathrm{max}}$=1060 is obtained, representing a fourfold improvement over the substrate-supported counterpart, and a fivefold improvement over the previous state of the art [27, 29]. Although the BIC is typically more sensitive to geometric perturbations than ordinary modes, $Q \geq 0.9Q_{\mathrm{max}}$ is maintained here over a lateral/height variation of approximately 15 nm, which is within the control of modern nanofabrication [38, 39].

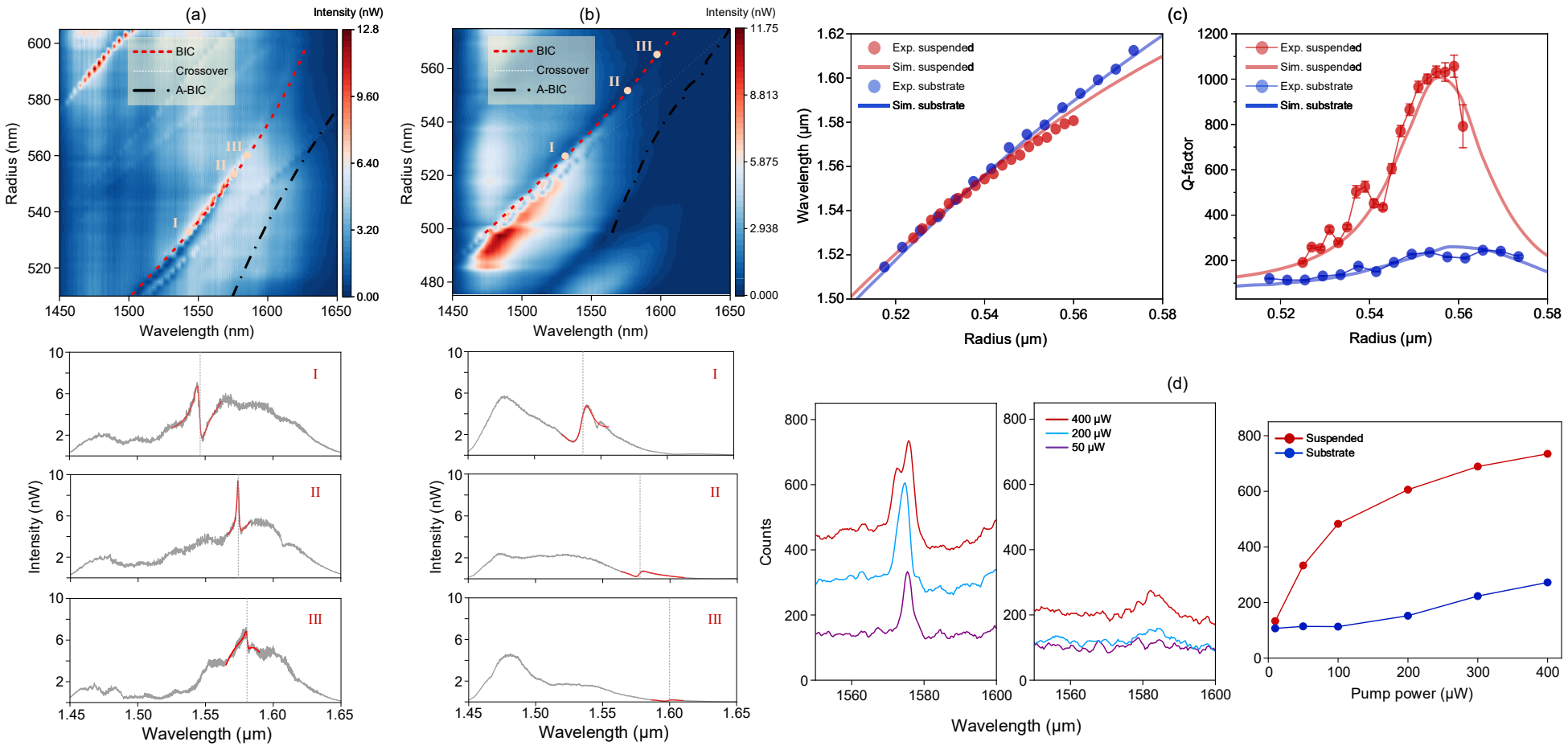

Fig. 4. Measured scattering spectra of InP nanocylinders suspended in air by nanobridges and designed to support O3 BIC-states. The spectra were acquired under azimuthally polarized excitation while varying the nanocylinder radius. The simulated BIC mode (red dashed line), crossover mode (white dotted line), and A-BIC mode (black dash-dotted line) are overlaid. The lower panels present representative scattering spectra for nanocylinders with radii corresponding to points I, II, and III in the 2D map shown above. The gray and red curves are the experimental result and Fano fitting, respectively, while the vertical dashed lines indicate the BIC positions. (b) Same as in (a), but for nanocylinders on a $SiO_2$ substrate. (c) Extracted wavelength (left) and $Q$-factor (right) of the BIC mode as functions of nanocylinder radius. Symbols and solid curves represent the experimental and FEM-simulated results, respectively. The red data correspond to suspended nanocylinders, while the blue data correspond to nanocylinders on the substrate. The $Q$-factor error bars indicate the standard deviation arising from multiple measurements and resonance fitting. The wavelength error bars are too small to be visible and are therefore omitted. (d) Photoluminescence spectra of suspended nanocylinders (left) and substrate-supported nanocylinders (middle) measured using a spectrometer at room temperature under 640 nm pulsed excitation (100 ps pulse width, 80 MHz repetition rate, 50x NA 0.65 objective) for varying average pump powers. The right panel plots the collected photoluminescence intensity at the BIC wavelength versus the pump power.

Practical applications, such as nonlinear optics and lasing, require not only strong optical confinement but also efficient excitation and signal collection. While our primary target applications are in-plane, on-chip photonic systems, valuable insight can be obtained from a more straightforward vertical characterization. We therefore performed photoluminescence (PL) measurements by exciting the structures with an above-bandgap pulsed laser (Fig. 4(d)). The suspended nanocylinders exhibit much stronger PL intensity than the substrate-supported ones. These results further confirm the importance of achieving high-$Q$ subwavelength cavities and highlight their potential for a wide range of applications.

## 5. Conclusion

We investigated BIC modes of different longitudinal orders in subwavelength nanocylinders while keeping the lowest azimuthal and transverse order, corresponding to one of the smallest footprints required to form a BIC. We identified an optimized aspect ratio and experimentally realized it in an InP nanocylinder, achieving a measured $Q$-factor exceeding 1000, in agreement with the design value and surpassing previous demonstrations. The advantages of the high-$Q$ resonance were demonstrated through both scattering and photoluminescence measurements. The device has a diameter of only ~1.1 μm ($0.7\lambda$) and operates at the telecom wavelength of 1.55 μm. This footprint approaches the layout area of CMOS standard cells, such as inverters and two-input logic gates, offering a promising platform for ultracompact light sources and nonlinear optical components integrated with electronics. In addition, our investigation provides an effective approach for assessing the limitations and potential of BIC modes in single subwavelength dielectric cavities, and can be extended to other materials and geometries, offering a possible route toward even higher-$Q$ optical devices with ultrasmall footprints.

**Funding.** This work was supported by the Villum Fonden via the Young Investigator Program (grant no. 42026 EXTREME), the Danish National Research Foundation (grant no. DNRF147 NanoPhoton), and the European Research Council (ERC) under the European Union Horizon 2020 Research and Innovation Programme (grant no. 834410 FANO).

**Acknowledgment.** The authors thank P. Holewa and E. Semenova for wafer epitaxy, and G. F. Sastre for measurement support.

**Disclosures.** The authors declare no conflicts of interest.

**Data Availability Statement.** Data underlying the results presented in this paper are not publicly available at this time but may be obtained from the authors upon reasonable request.

**Supplemental document.** See Supplement 1 for supporting content.

## Demonstration of a High-*Q* Subwavelength Dielectric Nanocylinder: Supplementary Document

Dayang Li[1,2], Simon Klinck Borregaard[1,2], Jesper Mørk[1,2], Meng Xiong[1,2], and Yi Yu[1,2,*]

[1]Department of Electrical and Photonics Engineering, Technical University of Denmark, Ørsteds Plads 345A, Kgs. Lyngby 2800, Denmark

[2]NanoPhoton - Center for Nanophotonics, Ørsteds Plads 345A, Kgs. Lyngby 2800, Denmark

*yiyu@dtu.dk

### Section 1. Estimation of the cylinder height and resonant frequency at mode intersection points

We consider a cylindrical cavity geometry supporting rotationally symmetric modes. In cylindrical coordinates ($\rho$, $\phi$, $z$), rotational symmetry implies the absence of azimuthal dependence. In this case, for the TE modes considered, $\mathbf{E} = E_\phi(\rho, z)\hat{\phi}$ is azimuthally polarized and the $E_\phi$ component serves as the pseudovector that fully describes the mode. The corresponding governing equation, derived from the Helmholtz equation, is given by [1]

$$\frac{\partial^2 E_\phi}{\partial z^2} + \frac{\partial^2 E_\phi}{\partial \rho^2} + \frac{1}{\rho}\frac{\partial E_\phi}{\partial \rho} - \frac{E_\phi}{\rho^2} + k^2 E_\phi = 0 \ , \tag{S.1}$$

where $k^2 = \mu\varepsilon\omega^2$ is the wavevector. Assuming a separable form $E_\phi(\rho, z) = R(\rho)Z(z)$, and substituting it into Eq. (S.1) we get

$$\frac{1}{Z}\frac{\partial^2 Z}{\partial z^2} + \frac{1}{R}\frac{\partial^2 R}{\partial \rho^2} + \frac{1}{\rho R}\frac{\partial R}{\partial \rho} - \frac{1}{\rho^2} + k^2 = 0 \ . \tag{S.2}$$

Now the $z$-dependent part and the $\rho$-dependent part must each equal a constant. Defining the separation constant as $-k_z^2$, we then have

$$\frac{1}{Z}\frac{\partial^2 Z}{\partial z^2} = -k_z^2 \Rightarrow \frac{\partial^2 Z}{\partial z^2} + k_z^2 Z = 0 \ , \tag{S.3}$$

with solutions $Z(z) = A\sin(k_z z) + B\cos(k_z z)$. Applying the hard boundary conditions at height $h$ gives the allowed axial wave numbers, i.e., $k_z = m\pi / h$. Subtracting Eq. (S.3) from Eq. (S.2), the radial equation becomes

$$\frac{\partial^2 R}{\partial \rho^2} + \frac{1}{\rho}\frac{\partial R}{\partial \rho} + \left(k^2 - k_z^2 - \frac{1}{\rho^2}\right)R = 0 \ . \tag{S.4}$$

Defining $\beta^2 = k^2 - k_z^2$ and multiplying Eq. (S.4) by $\rho^2$ leads to

$$\rho^2 R'' + \rho R' + \left(\beta^2\rho^2 - 1\right)R = 0 \ . \tag{S.5}$$

This is the Bessel equation of order 1, so the solution is $R(\rho)=J_1(\beta\rho)$. Now impose the hard boundary conditions at the cylinder sidewall $\rho=r$, we get $R(r)=0 \Rightarrow J_1(\beta r)=0$. So $\beta r$ must be the $n$-th zero of $J_1$, denoted as $x_n$. Therefore $\beta = x_n / r$. The, the full solution of $E_\phi(\rho,z)$ becomes

$$E_\phi(\rho,z)^{(n,m)} = J_1\left(\frac{x_n}{r}\rho\right)\left(A_{nm}\sin\left(\frac{m\pi}{h}z\right)+B_{nm}\cos\left(\frac{m\pi}{h}z\right)\right),$$ ($A_{nm}$=0 for odd $m$, and $B_{nm}$=0 for even $m$).

(S.6)

At the same time, since $\beta^2 = k^2 - k_z^2$, we have $k^2 = (x_n / r)^2 + k_z^2 \Rightarrow \omega\sqrt{\mu\varepsilon} = \sqrt{(x_n / r)^2 + k_z^2}$. So

$$\omega_{m,n} = \frac{1}{\sqrt{\mu\varepsilon}}\sqrt{(x_n / r)^2 + k_z^2} \quad . \tag{S.7}$$

At the crossing point between the FP-like mode ($m$=O+2, $n$=1) and the Mie-like mode ($m$=O, $n$=2), where O=1,2,3…, Eq. (S.7) yields the nanocylinder height $h_\text{O}$ as

$$\begin{gathered}(x_1 / r)^2 + \frac{(\text{O}+2)^2}{h_\text{O}^2}\pi^2 = (x_2 / r)^2 + \frac{\text{O}^2}{h_\text{O}^2}\pi^2 \Rightarrow \\ h_\text{O} = 2\pi r\sqrt{(\text{O}+1)/(x_2^2 - x_1^2)}\end{gathered} \quad . \tag{S.8}$$

Substituting $h_\text{O}$ gives

$$\omega_0 \approx \omega_{\text{O}+2,1} = \omega_{\text{O},2} = \frac{c}{2n_e r}\sqrt{\frac{(\text{O}+2)^2 x_2^2 - \text{O}^2 x_1^2}{\text{O}+1}} \quad . \tag{S.9}$$

Using $r$=0.56 μm, $n_e$=3.6 (an approximately value for O≥3 for our InP nanocylinders), we have $\omega_0 \approx 4.37\sqrt{\text{O}+4.7}\times 10^{14}$ rad/s.

**Section 2. Coupling coefficients extracted from quasinormal-mode perturbation theory**

The perturbed modes $|\mathbf{a}\rangle$ and their eigenfrequencies $\tilde{\omega}_\text{PT}$ can be estimated from the unperturbed modes using quasinormal-mode perturbation theory (QNM-PT), by solving the following eigenvalue equation [2]

$$\mathbf{H}_0|\mathbf{a}\rangle = \tilde{\omega}_\text{PT}(\mathbf{I}+\mathbf{H}_\text{P})|\mathbf{a}\rangle \Rightarrow \mathbf{H}_\text{PT}|\mathbf{a}\rangle = \tilde{\omega}_\text{PT}|\mathbf{a}\rangle \ , \tag{S.10}$$

with

$$\mathbf{H}_\text{PT} = [\mathbf{I}+\mathbf{H}_\text{P}]^{-1}\mathbf{H}_0 \ . \tag{S.11}$$

Here, $\mathbf{H}_0$ and $\mathbf{H}_\text{P}$ are given by Eq. (3) in the main text. If we focus on two-mode interactions, which is typically the scenario for BIC formation (only two modes are found to exhibit non-negligible off-diagonal coupling), we can extract a reduced 2×2 submatrix from $\mathbf{H}_\text{PT}$ as

$$\mathbf{H}_{\text{PT},2} = \begin{bmatrix} a & b \\ c & d \end{bmatrix} . \tag{S.12}$$

The off-diagonal elements in $\mathbf{H}_{\mathrm{PT},2}$ are generally asymmetric ($b \neq c$). This asymmetry does not affect the eigenvalues or the underlying modal interaction, but arises as a representation effect when converting the generalized QNM-PT problem into a standard eigenvalue problem (Eq. (S.10)). It therefore should not be interpreted as nonreciprocal coupling or as different coupling strengths in opposite directions. To obtain a symmetric representation, we apply a diagonal similarity transformation.

$$\mathbf{H}'_{\mathrm{PT},2} = \mathbf{S}^{-1}\mathbf{H}_{\mathrm{PT},2}\mathbf{S},$$

where $\mathbf{S} = \mathrm{diag}\left((b/c)^{1/4}, (c/b)^{1/4}\right)$, and we thus obtain

$$\mathbf{H}'_{\mathrm{PT},2} = \begin{bmatrix} a & \sqrt{bc} \\ \sqrt{bc} & d \end{bmatrix}, \tag{S.13}$$

which has the same eigenvalues as $\mathbf{H}_{\mathrm{PT},2}$, but with rescaled modal amplitudes $|\mathbf{a}\rangle' = \mathbf{S}^{-1}|\mathbf{a}\rangle$. The off-diagonal term $\eta = \sqrt{bc} = \eta_r + j\eta_i$ represents the coupling between the FP- and Mie-like modes. It also reflects how strongly the geometric perturbation modifies the destructive interference between the two modes, thereby determining the $Q$-factor limit.

The individual matrix elements of $\mathbf{H}'_{\mathrm{PT},2}$ are not directly interpretable as standard CMT parameters [3]-[5]. For example, the complex diagonal elements $a$ and $d$ of $\mathbf{H}'_{\mathrm{PT},2}$ may, in some cases, have positive imaginary parts (even though the eigenvalues of $\mathbf{H}'_{\mathrm{PT},2}$ still retain negative imaginary parts), which would imply an unphysical effective gain. We therefore construct an equivalent two-mode CMT model, $\mathbf{H}_{\mathrm{CMT}}$, as given in Eq. (4) of the main text, which preserves the same trace and determinant, and hence the same eigenvalues

$$\mathrm{tr}\left(\mathbf{H}'_{\mathrm{PT},2}\right) = \mathrm{tr}\left(\mathbf{H}_{\mathrm{CMT}}\right) = \mathrm{T},\ \det\left(\mathbf{H}'_{\mathrm{PT},2}\right) = \det\left(\mathbf{H}_{\mathrm{CMT}}\right) = \mathrm{D}. \tag{S.14}$$

Since this mapping is not unique, we consider two representative coupled systems commonly used to interpret BIC formation. In these systems, destructive interference arises when two modes possess similar far-field radiation patterns but radiate out of phase.

1) In the first model (CMT1), we impose

$$\mathbf{H}_{\mathrm{CMT}} = \begin{bmatrix} \omega_1 - j\gamma & \kappa_r + j\gamma \\ \kappa_r + j\gamma & \omega_2 - j\gamma \end{bmatrix}, \tag{S.15}$$

so $\gamma_1 = \gamma_2 = \gamma$ and $\kappa_i = \sqrt{\gamma_1\gamma_2} = \gamma$, meaning that the loss coupling is completely through each mode's own radiation. Then, using the criteria of Eq. (S.14), we get

$$\begin{aligned} &\gamma = -\mathrm{Im}\{\mathrm{T}\}/2 \\ &\kappa_r = \left(2\,\mathrm{Im}\{\mathrm{D}\}/\mathrm{Im}\{\mathrm{T}\} - \mathrm{Re}\{\mathrm{T}\}\right)/2 \\ &\omega_{1,2} = \left(\mathrm{Re}\{\mathrm{T}\} \pm \Delta\omega\right)/2,\ \Delta\omega = \sqrt{\mathrm{Re}\{\mathrm{T}\}^2 - 4\left(\mathrm{Re}\{\mathrm{D}\} + \kappa_r^2\right)} \end{aligned} \tag{S.16}$$

2) In the second model (CMT2), we impose

$$\mathbf{H}_{\mathrm{CMT}}=\begin{bmatrix}\omega_0-j\gamma & \kappa_r+j\kappa_i\\ \kappa_r+j\kappa_i & \omega_0-j\gamma\end{bmatrix},\tag{S.17}$$

so $\omega_1=\omega_2=\omega_0$ and $\gamma_1=\gamma_2=\gamma$, meaning that the uncoupled modes are degenerate. In this case, $|\kappa_i|<\gamma$, which is general. Then, using the criteria of Eq. (S.14), we get

$$\begin{aligned}&\omega_0=\mathrm{Re}\{\omega_++\omega_-\}/2\\ &\gamma=-\mathrm{Im}\{\omega_++\omega_-\}/2\ ,\\ &\kappa=(\omega_+-\omega_-)/2\end{aligned}\tag{S.18}$$

where $\omega_\pm$ are the eigenvalues of $\mathbf{H}'_{\mathrm{PT},2}$.

## Section 3. Analysis based on coupled mode theory

### A. *Q*-factor of the BIC mode

For the Friedrich-Wintgen BIC, the standard CMT [5] description involves two resonances residing in the same cavity that are coupled through the radiation channel

$$\frac{da_1}{dt}=-j\omega_1a_1-\gamma_1a_1-j(\kappa_r+j\kappa_i)a_2\ ,\tag{S.19}$$

$$\frac{da_2}{dt}=-j\omega_2a_2-\gamma_2a_2-j(\kappa_r+j\kappa_i)a_1\ .\tag{S.20}$$

Here, $a_i$ ($i=1,2$) denote the field amplitudes. Eqs. (S.19) and (S.20) yield the eigenmodes of the system with the eigen frequencies

$$\omega_\pm=\frac{\omega_1+\omega_2}{2}-j\frac{\gamma_1+\gamma_2}{2}\pm\sqrt{Z}\ ,\ \text{with}\ Z=\left(\frac{\Delta\omega}{2}-j\frac{\Delta\gamma}{2}\right)^2+(\kappa_r+j\kappa_i)^2.\tag{S.21}$$

Here we define $\Delta\omega=\omega_1-\omega_2$ and $\Delta\gamma=\gamma_1-\gamma_2$. Since the BIC of the nanocylinder has a higher resonant frequency and *Q*-factor than the A-BIC, and if we impose $\gamma_1=\gamma_2=\gamma$, as shown in Section 2, it follows, after some algebra, that

$$\begin{aligned}&\mathrm{Re}\{\omega_{\mathrm{BIC}}\}=\mathrm{Re}\{\omega_+\}=\frac{\omega_1+\omega_2}{2}+u\\ &\mathrm{Im}\{\omega_{\mathrm{BIC}}\}=\mathrm{Im}\{\omega_+\}=-\frac{\gamma_1+\gamma_2}{2}+\upsilon\end{aligned}\ ,\tag{S.22}$$

where

$$u=\sqrt{\frac{1}{2}\left(\sqrt{\left(\frac{\Delta\omega^2}{4}+\left(\kappa_r^2-\kappa_i^2\right)\right)^2+\left(2\kappa_r\kappa_i\right)^2}+\frac{\Delta\omega^2}{4}+\left(\kappa_r^2-\kappa_i^2\right)\right)}\ ,\tag{S.23}$$

and

$$\upsilon=\sqrt{\frac{1}{2}\left(\sqrt{\left(\frac{\Delta\omega^2}{4}+\left(\kappa_r^2-\kappa_i^2\right)\right)^2+\left(2\kappa_r\kappa_i\right)^2}-\frac{\Delta\omega^2}{4}-\left(\kappa_r^2-\kappa_i^2\right)\right)}\ . \tag{S.24}$$

In this CMT1 case, one has $\Delta\omega^2/4<<|\kappa|^2$ (typically valid near the avoided crossing point), and $\kappa_i=\gamma$, the BIC's $Q$-factor can be approximated as (see Section 3.B)

$$Q_{\rm BIC}=\frac{\mathrm{Re}\{\omega_{\rm BIC}\}}{-2\,\mathrm{Im}\{\omega_{\rm BIC}\}}\approx\frac{\omega_0}{2\left(\Delta\omega^2|\kappa_i|/\left(8\left(\kappa_r^2+\kappa_i^2\right)\right)\right)}=Q_m\left(8|\kappa|^2\right)/\Delta\omega^2\ , \tag{S.25}$$

where $\omega_0=(\omega_1+\omega_2)/2$ and $Q_m=\omega_0/(2\gamma)$.

In the CMT2 case where $\Delta\omega=0$, we have $u=|\kappa_r|$, $\upsilon=|\kappa_i|$, and thus

$$Q_{\rm BIC}=\frac{\mathrm{Re}\{\omega_{\rm BIC}\}}{-2\,\mathrm{Im}\{\omega_{\rm BIC}\}}\approx\frac{\omega_0}{2\left(\gamma-|\kappa_i|\right)}=Q_m/\left(1-|\kappa_i|/\gamma\right)\ . \tag{S.26}$$

**B. Proof that $u$ is dominated by $\kappa_r$ and $\upsilon$ is dominated by $\kappa_i$ at the avoided crossing point**

For $\Delta\gamma=0$, we have from Eq. (S.21)

$$Z=\left(\kappa_r+j\kappa_i\right)^2+\left(\frac{\Delta\omega}{2}\right)^2=\kappa_r^2-\kappa_i^2+\frac{\Delta\omega^2}{4}+2j\kappa_r\kappa_i\,.$$

When $|\Delta\omega|$ is small (typical case at the avoided crossing point), i.e., $\Delta\omega^2/4<<|\kappa|^2=\kappa_r^2+\kappa_i^2$, we can make a Tylor expansion at $Z_0=\left(\kappa_r+j\kappa_i\right)^2$, so that

$$\sqrt{Z}\approx\sqrt{Z_0}+\frac{1}{2}\frac{\Delta Z}{\sqrt{Z_0}}=\left(1+\frac{\Delta\omega^2}{8\left(\kappa_r^2+\kappa_i^2\right)}\right)\kappa_r+j\left(1-\frac{\Delta\omega^2}{8\left(\kappa_r^2+\kappa_i^2\right)}\right)\kappa_i\ . \tag{S.27}$$

So its real and imaginary parts become

$$u\approx\left(1+\frac{\Delta\omega^2}{8\left(\kappa_r^2+\kappa_i^2\right)}\right)\kappa_r\ ,\ \upsilon\approx\left(1-\frac{\Delta\omega^2}{8\left(\kappa_r^2+\kappa_i^2\right)}\right)\kappa_i\ . \tag{S.28}$$

Eqs. (S.28) shows that the correction to $\kappa_r$ and $\kappa_i$ is only second order in $\Delta\omega$. That means $u$ ($\upsilon$) is largely set by $\kappa_r$ ($\kappa_i$), unless the frequency mismatch becomes large.

**Section 4. BIC versus whispering gallery modes**

Fig. S1 compares the $Q$-factors of the BIC (red curves) and whispering gallery modes (blue curves). The whispering-gallery modes included here are those with $Q$-factors higher than the maximum BIC-$Q$ for each order and wavelengths within a 100 nm window centered at the optimum BIC wavelength of that order.

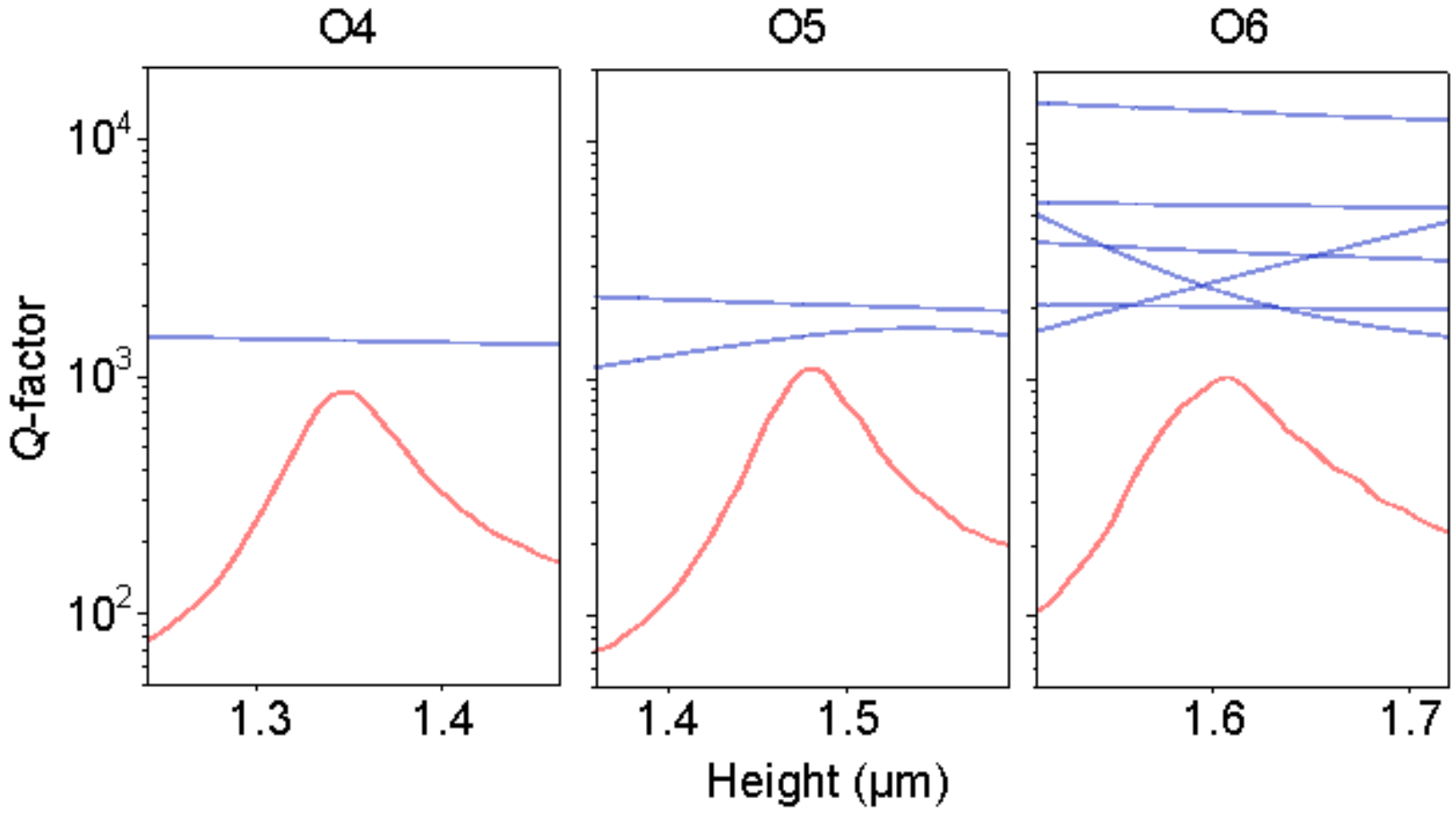

**Fig. S1.** $Q$-factors of the BIC modes (red) and whispering gallery modes (blue) as functions of nanocylinder height. The left, middle, and right panels correspond to orders O4, O5, and O6, respectively.

**Section 5. Mode fields in nanocylinders supported by nanobridges**

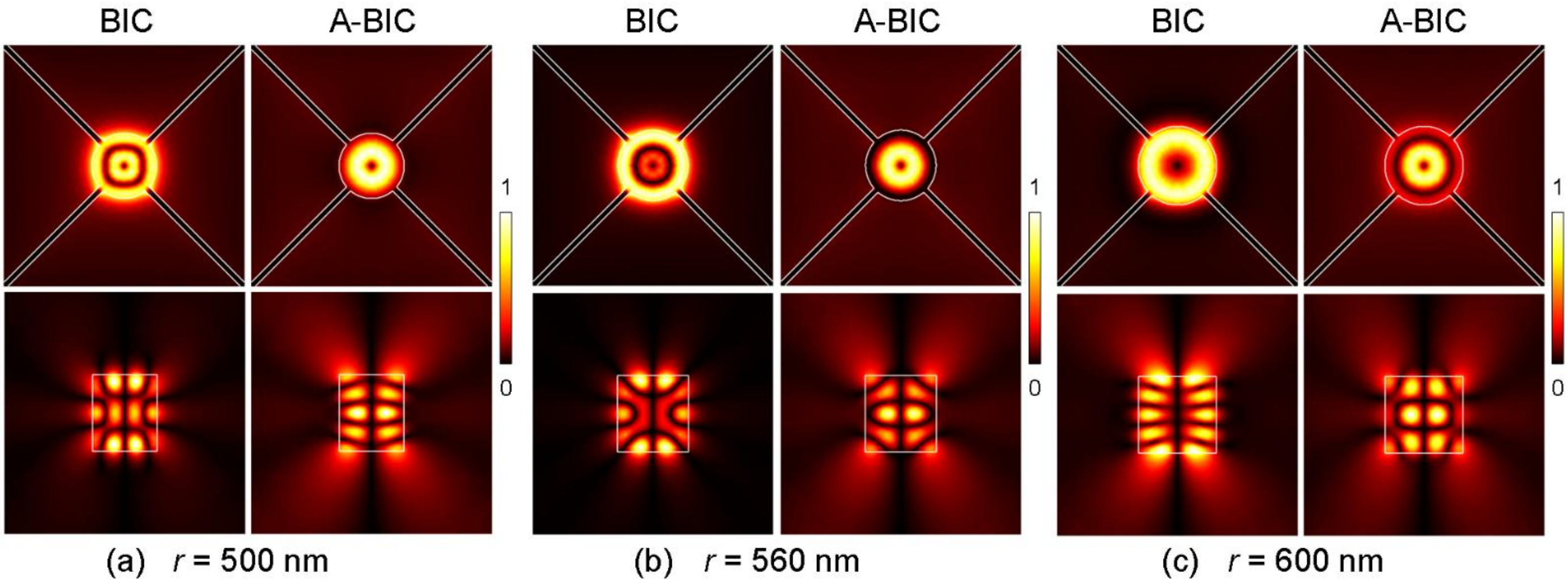

**Fig. S2.** Mode-field profiles (|**E**|) of InP nanocylinders with bridges, shown in the central transverse $x$-$y$ planes (top row) and vertical $x$-$z$ cross sections (bottom row). The cylinder radii are (a) 500 nm, (b) 560 nm, and (c) 600 nm. The left panels show the BIC mode branch, while the right panels show the A-BIC mode branch.

Fig. S2 shows the mode profiles of the InP nanocylinders with supporting bridges. As the radius passes through the optimum BIC point, corresponding to the avoided crossing at $r$=560 nm, the mode pattern of the BIC (at $r$=500 nm) evolves into that of the A-BIC (at $r$=600 nm), and vice versa. This reflects the exchange of modal character between the FP-like and Mie-like branches. The introduction of nanobridges to suspend the cylinder causes almost no perturbations to the original modes, with no additional leakage and only a negligible effect on the eigenfrequencies.

**Section 6. Scattering measurement with radial polarization**

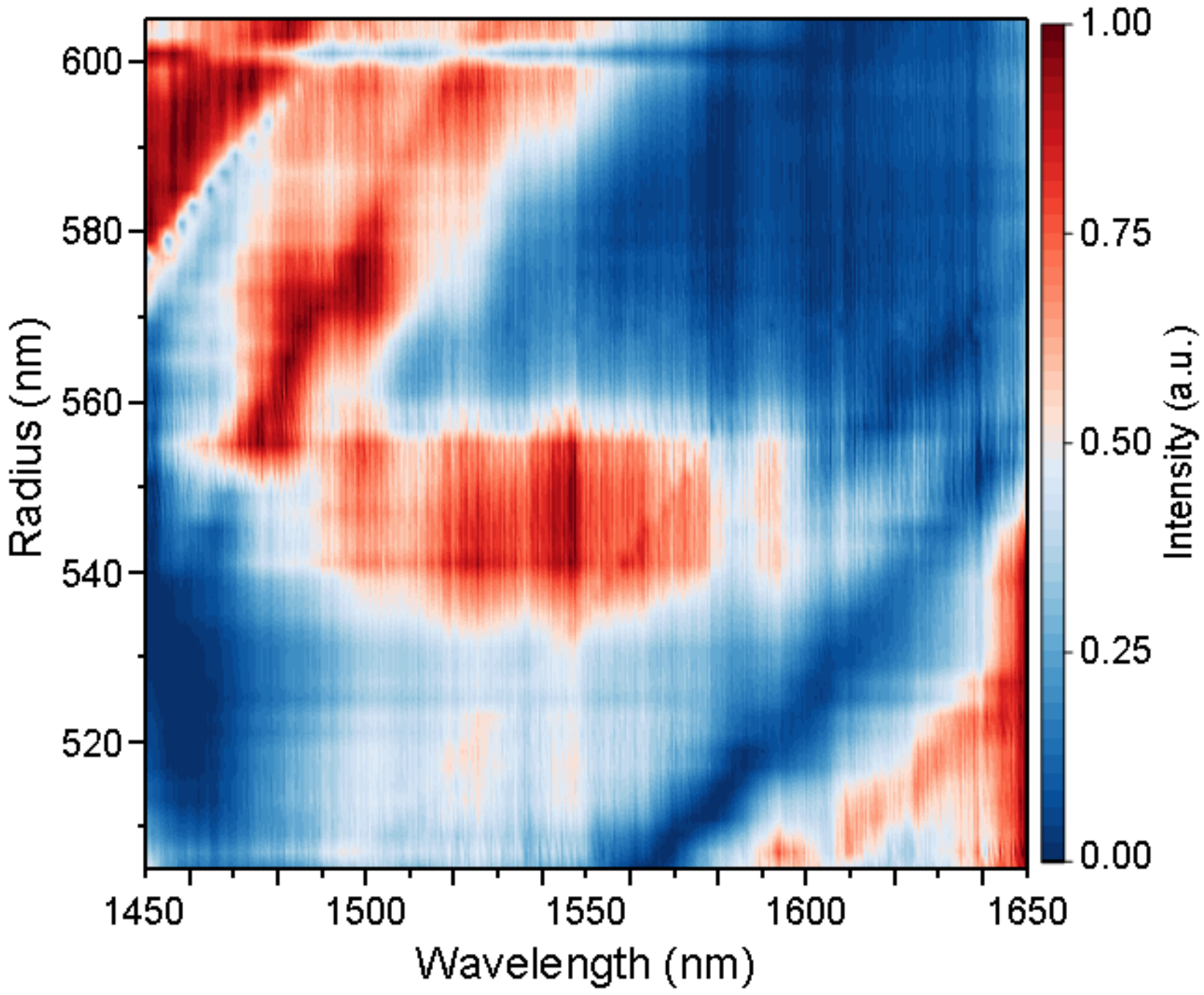

**Fig. S3.** Measured scattering spectra for InP O3 nanocylinders with radii ranging from 500 to 600 nm under radially polarized excitation.

As shown in Fig. S3, under radially polarized excitation, the BIC mode (indicated by the red curve in Fig. 4(a) in the main text) becomes nearly invisible due to inefficient excitation, as the mode is dominated by the $E_\phi$ component. The weak residual scattering likely arises from incomplete conversion of the incident linearly polarized light into radial polarization, leaving a small azimuthally polarized component. This is because our broadband SLED source is not uniformly covered by the operating wavelength ranges of the polarization optics, such as the Faraday rotator and q-plate.